\documentclass[
  reprint,
  superscriptaddress, 
  amsmath,amssymb,
  aps,
  prl
]{revtex4-2}

\usepackage{graphicx}
\usepackage{dcolumn}
\usepackage{bm}
\usepackage{dsfont}
\usepackage{tikz}
\usepackage{indentfirst}
\usepackage{multirow}

\usepackage[colorlinks,citecolor=blue,linkcolor=red,urlcolor=magenta]{hyperref}

\newcommand{\supplementarysection}{%
  \setcounter{figure}{0}
  \let\oldthefigure\thefigure
  \renewcommand{\thefigure}{S\oldthefigure}
  \setcounter{section}{0}
  \let\oldthesection\thesection
  \renewcommand{\thesection}{S\oldthesection}
  \setcounter{equation}{0}
  \let\oldtheequation\theequation
  \renewcommand{\theequation}{S\oldtheequation}
  \setcounter{table}{0}
  \let\oldthetable\thetable
  \renewcommand{\thetable}{S\oldthetable}
}
\newcommand{\remove}[1]{}

\newcommand{\bi}{\begin{itemize}}
\newcommand{\ei}{\end{itemize}}
\newcommand{\be}{\begin{enumerate}}
\newcommand{\ee}{\end{enumerate}}
\newenvironment{dfn}{{\vspace*{1ex} \noindent \bf Definition }}{\vspace*{1ex}}

\newcommand{\nn}{\nonumber}  %

\newcommand{\Eq}[1]{Eq.~(\ref{#1})}

\newcommand{\trace}[1]{\mathrm{tr}{\left(#1\right)}}
	
	\newcommand{\beq}{\begin{eqnarray}}
	\newcommand{\eeq}{\end{eqnarray}}
	\newcommand{\bea}{\begin{eqnarray}\begin{aligned}}
	\newcommand{\eea}{\end{aligned}\end{eqnarray}}

\definecolor{forestgreen}{RGB}{34,139,34}

\begin{document}
\title{Topological Charge-$2ne$ Superconductivity and Clustering Hierarchy}
\author{Zhi-Qiang Gao}
\thanks{These authors contributed equally}
\email{zqgao@berkeley.edu}
\affiliation{Department of Physics, University of California, Berkeley, California 94720, USA}
\author{Yan-Qi Wang}
\thanks{These authors contributed equally}
\email{wyq@umd.edu}
\affiliation{Department of Physics and Joint Quantum Institute, University of Maryland,
College Park, Maryland 20742, USA}
\author{Hui Yang}
\thanks{These authors contributed equally}
\email{huiyang.physics@gmail.com}
\affiliation{Department of Physics and Astronomy, Johns Hopkins University, Baltimore, Maryland 21218, USA}
\affiliation{Department of Physics and Astronomy, University of Pittsburgh, Pennsylvania 15213, USA}
\author{Congjun Wu}
\email{wucongjun@westlake.edu.cn}
\affiliation{New Cornerstone Science Laboratory, Department of Physics, School of Science, Westlake University, Hangzhou 310024, Zhejiang, China}
\affiliation{Institute of Natural Sciences, Westlake Institute for Advanced Study, Hangzhou 310024, Zhejiang, China}
\affiliation{Institute for Theoretical Sciences, Westlake University, Hangzhou 310024, Zhejiang, China}
\affiliation{Key Laboratory for Quantum Materials of Zhejiang Province, School of Science, Westlake University, Hangzhou 310024, Zhejiang, China}

\begin{abstract}
Charge-$2ne$ superconductors condense clusters of more than two electrons, extending superconductivity beyond Cooper pairing. They exhibit distinctive signatures including fractional flux quantization $h/(2ne)$ and are actively being explored in strongly correlated systems. We develop a general framework for constructing and characterizing \emph{topological} charge-$2ne$ superconductors by combining wavefunction and field theory approaches. Generalizing the Read--Green wavefunction and Higgsing Read--Rezayi-type states yield two families that can support deconfined non-Abelian anyons and non-Abelian superconducting vortices. The first family also admits a construction from doping Jain states. Through Higgs theories, we find that the first and second families realize weak and intermediate clustering regimes, respectively, distinguished by the topology of vortices carrying flux greater than $h/(2ne)$, while strong clustering yields a charge-$2ne$ superconductor without intrinsic topological order. These reveal that higher-charge superconductivity supports a richer hierarchy of topological phases beyond the Read--Green weak--strong pairing dichotomy, with distinct internal clustering structures giving rise to different topological structures at fixed condensate charge.
\end{abstract}


\maketitle


{\it Introduction.---} Superconductivity usually arises from condensation of charge-$2e$ Cooper pairs. However, this is not the only way a superconducting state can form. The elementary condensed object can be composed of more than two electrons. 
For example, in a charge-$4e$ superconductor, electron quartets condense while Cooper pairs remain gapped. Quartetting has been investigated in anyon superconductivity~\cite{Lee1989,Pichler2025}, $SU(4)$ symmetric fermionic systems~\cite{Wu2005,Lecheminant2005,Capponi2007,Capponi2008,Roux2008,Ko2008,Zhang2020}, moir\'e systems~\cite{Khalaf2022,Wu2024,Kim2025}, and doped symmetric Mott insulators~\cite{Zhang20254e,Gao20264e}. It can also emerge as vestigial order in pair density wave and multicomponent superconductors~\cite{Berg2009,Radzihovsky2009,Radzihovsky2011,Babaev2004,Han2020,Jian2021,Fernandes2021,Grinenko2021,Gao2022,Jin2022,Han2026}. Fractional $h/(4e)$ flux quantization carried by charge-$4e$ SC vortices provides an experimental diagnostic~\cite{Gladchenko2008,Ge2024,Lin2025}.

The interplay with topology raises a distinct question. Read and Green connected the weak-pairing $(p+ip)$-wave superconductor to the Moore--Read quantum Hall state using a wavefunction approach, revealing Majorana vortex modes and a chiral Majorana edge~\cite{Moore1991,Read2000,Nayak2008,ZhangRMP,Sato2017}. Their weak--strong pairing distinction illustrates how phases with identical superconducting order can have different topology. Higher-charge superconductivity, however, lies beyond a quadratic mean-field description. What topological structures it can support and how the internal organization of an electronic cluster distinguishes different superconducting phases remain unclear. The proximity of superconductivity and fractionalized topological phases in correlated moir\'e systems further motivates these questions~\cite{Cai2023FQAH,Han2024signatures,Chou2024intravalley,Kim2025,Geier2024Chiral,Wang2024,maymann2025,Xu2025FQAHSC}.

In this Letter, we construct and characterize topological charge-$2ne$ superconductors using wavefunction-inspired conformal field theories (CFTs) and bulk topological quantum field theories (TQFTs)~\cite{Moore1991,Barkeshli2010,Barkeshli2010parton}. In two dimensions, intrinsic topological order supports deconfined anyons with nontrivial fusion and braiding~\cite{WenRMP}. For the chiral phases considered here, Chern--Simons (CS) theories encode these topological data, which are related to edge CFTs through the bulk--edge correspondence~\cite{Wen1994,Wen1995,Qi2006,Witten2016}. Identifying the physical electron then distinguishes deconfined anyons from superconducting vortex sectors. We treat the gapless superconducting phase mode separately from the gapped topological sector. 

We first generalize the Read--Green wavefunction to $2n$-electron clusters, obtaining topological charge-$2ne$ superconductors with deconfined non-Abelian bulk anyons and non-Abelian vortices. A second construction starts from Read--Rezayi quantum Hall wavefunctions~\cite{Read1999}, which exhibit $2n$-cluster structures of particles coupled to dynamical $U(1)$ gauge fields, and Higgses these gauge fields to generate $2n$-clustering of physical electrons, producing a distinct family of topological charge-$2ne$ superconductors~\cite{Wang2024the,Pace2024}. Although they share the same condensate charge $2ne$, these families differ in the topology of vortices carrying flux greater than $h/(2ne)$ and the organization of their constituent clusters. Crucially, a Higgs theory connects the first family to representative members of the second family as weak- and intermediate-clustering phases, respectively, and relates them to a topologically trivial~\footnote{Here ``topologically trivial'' means the absence of intrinsic topological order. The strong-clustering endpoint may still differ from a molecular Bose--Einstein condensate by an invertible layer; e.g., for $n=1$, if $U(2)_{2,0}$ is completed by a decoupled $U(1)_{-1}$ layer to describe the $(p+ip)$-wave superconductor, the endpoint carries $c_-=-1$.} strong-clustering limit. The resulting hierarchy demonstrates how charge-$2ne$ superconductors support interacting topological phases of matter beyond the conventional Bogoliubov--de Gennes paradigm.

\begin{figure}[htbp]
  \centering
  \includegraphics[width=0.55\linewidth]{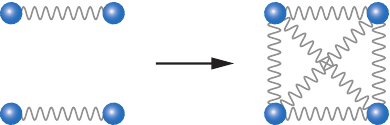}
  \caption{Illustration of generalizing the Read--Green wavefunction using the charge-$4e$ case as an example. The blue dots represent electrons. Each gray wavy segment represents a $(z-z^\prime)^{-1}$ factor from $(p+ip)$-wave pairing.}
  \label{fig:quartet}
\end{figure}

{\it Generalizing the Read--Green wavefunction.---}
We first consider building up the topological charge-$2ne$ superconductor from Read--Green ingredients. The Read--Green Pfaffian wavefunction for the $(p+ip)$-wave topological superconductor is built from the two-point real space form factor $(z_1-z_2)^{-1}$, which corresponds to the $(p+ip)$-wave pairing in $k$-space. For the charge-$2ne$ superconductor, a direct extension is simply to introduce pairing between each pair of electrons within the $2n$-electron cluster. The elementary wavefunction $F$ is then
\beq\label{eq:elementary_wavefunction}
F(z_1,\cdots,z_{2n})=\prod_{1\le i<j\le 2n}\frac{1}{z_i-z_j},
\eeq
as illustrated in Fig.~\ref{fig:quartet}.

According to the bulk--edge correspondence, the edge CFT should have the $2n$-point function of the edge electron operator $\psi(z)$ that exactly produces the elementary wavefunction, as given in \Eq{eq:elementary_wavefunction}, $\big<\prod_{i}\psi(z_i)\big>=F(z_1,\cdots,z_{2n})$. In addition, such a CFT must satisfy the constraints that (i) fusion results of $q<2n$ electron operators cannot contain the identity operator to prevent the charge-$qe$ instability, and (ii) the electron operator should have half-odd integer conformal weight and be identified as the transparent fermion~\footnote{In fact, these two conditions are both consequences of $\big<\prod_{i}\psi(z_i)\big>=F(z_1,\cdots,z_{2n})$, and the conformal weight of the electron operator is fixed to $n-1/2$.}. The simplest choice is the $SU(2n)_{2n}$ Wess--Zumino--Witten CFT, where the fermionic primary field $\Lambda_1(z)$ labeled by $SU(2n)$ weight $(2n,0,\cdots,0)$ is identified as the electron $\psi(z)$, and the many-electron state is defined by the correlator $\langle\prod_i\Lambda_1(z_i)\rangle$~\footnote{For more than $2n$ electrons, the correlator $\langle\prod_i\Lambda_1(z_i)\rangle$ is not an antisymmetrized product of $F$'s. We define the state by this correlator, whose single-cluster amplitude is $F$; the Moore--Read state is likewise defined by a conformal-block correlator.}. This identification converts the theory into the fermionic $SU(2n)_{2n}/\mathbb{Z}_{2n}$ spin CFT~\footnote{Since $\Lambda_1$ generates a one-form $\mathbb{Z}_{2n}$ center symmetry in the $SU(2n)_{2n}$ CFT, identifying it with the physical electron (transparent fermion) gauges the one-form $\mathbb{Z}_{2n}$ symmetry.}. The corresponding bulk theory is the $U(2n)_{2n,0}$ fermionic CS theory with chiral central charge $c_-=2n^2-1/2$:
\beq
\mathcal{L}_{2ne}^\mathrm{CS}[A]&=&\frac{2n}{4\pi}\trace{a\mathrm{d}a+\frac{2}{3}a^3}-\frac{1}{4\pi}(\mathrm{tr}\,a)\mathrm{d}(\mathrm{tr}\,a)\nn\\
&&-\frac{1}{2\pi}(\mathrm{tr}\,a)\mathrm{d}A,\label{eq:LCS}
\eeq
where $a$ is the $U(2n)$ gauge field, and $A$ is the background electromagnetic field. By taking $a=\alpha\mathbf{1}_{2n}$, it can be easily seen that \Eq{eq:LCS} indeed describes a charge-$2ne$ superconductor. In particular, for $n=1$, the theory is reduced to $U(2)_{2,0}$, which describes the $(p+ip)$-wave topological superconductor~\footnote{Strictly speaking, one needs to further stack a decoupled $U(1)_{-1}$. See also Refs.~\cite{Ma2020,Shi2025}.}. The topological data of this series of charge-$2ne$ superconductors can be deduced via a standard anyon condensation procedure~\cite{Burnell2018}. Here we list the anyon content of the bulk theory for the $n=2$ case in Table~\ref{tab:2}, where $\sigma_\pm$ realize the ``silver ratio'' quantum dimension $1\!+\!\sqrt{2}$ carried by the nontrivial anyon in $SO(3)_3$ fermionic CS theory~\cite{Fidkowski2013} and $\epsilon=\sigma_+\times\sigma_-$ is their composite. This suggests an $SO(3)_3\otimes SO(3)_3$ bulk topological order. The superconducting vortices in this state are also non-Abelian. The $h/(4e)$ elementary vortices have quantum dimensions $\sqrt{4\!+\!2\sqrt{2}}$ and $(1\!+\!\sqrt{2})\sqrt{4\!+\!2\sqrt{2}}$, while the $h/(2e)$ ones have $2\!+\!\sqrt{2}$ and $2\!+\!2\sqrt{2}$. We give a general recipe of construction from anyon condensation in the Supplemental Material~\cite{supp}.

\begin{table}[!htbp]
\begin{tabular}{ccccc}
\hline\hline
Anyon type & $1$ & $\epsilon$ & $\sigma_+$ & $\sigma_-$\\
\hline
Topo. spin & $0$ & $1/2$ & $1/4$ & $1/4$ \\
Quantum dim. & $1$ & $3+2\sqrt{2}$ & $1+\sqrt{2}$ & $1+\sqrt{2}$\\
\hline\hline
\end{tabular}
\caption{Anyon content for the topological charge-$4e$ superconductor constructed from generalizing the Read--Green wavefunction. Because this state is a fermionic topological order, the topological spin of an anyon is defined modulo $1/2$.}\label{tab:2}
\end{table}

The $U(2n)_{2n,0}$ topological charge-$2ne$ superconductor also admits a construction from doping a Jain state at $\nu=2n/(2n-1)^2$, which provides a complementary perspective on the formation of such a state. We illustrate this construction for $n=2$, leaving the generalizations to the Supplemental Material~\cite{supp}.

The parent $\nu=4/9$ Jain state admits a $U(4)_{1,36}$ CS description. We assume that its lowest-energy charged excitation is the elementary anyon carrying charge $e/9$, such that doping the state introduces a finite density of these anyons. The resulting low-energy theory is
\beq
\mathcal{L}_\mathrm{Jain~4/9}^\mathrm{doped}[A]&=&\frac{1}{4\pi}\trace{a\mathrm{d}a+\frac{2}{3}a^3}+\frac{2}{4\pi}(\mathrm{tr}\,a)\mathrm{d}(\mathrm{tr}\,a)\nn\\
&&-\frac{1}{2\pi}(\mathrm{tr}\,a)\mathrm{d}A+\mathcal{L}[\phi,a],\label{eq:Jain}
\eeq
where the bosonic matter field $\phi$ sources the elementary anyon and transforms in the fundamental representation $(\mathbf{4},1)$ of $U(4)$. Its dynamics and coupling to the $U(4)$ gauge field $a$ are encoded in $\mathcal{L}[\phi,a]$. For the lattice realization of a Jain-$4/9$ fractional quantum anomalous Hall insulator, the translations act projectively on the elementary anyon as $T_xT_y=e^{2\pi i/9}T_yT_x$. This organizes the low-energy modes of $\phi$ into nine symmetry-related valleys. The corresponding fields are $\phi_{\sigma,v}$, where $\sigma=1,2,3,4$ labels the $U(4)$ gauge indices and $v=1,2,\ldots,9$ labels the valleys. An ansatz invariant under both $U(4)$ and the magnetic translations has equal density of these components and no average non-Abelian gauge flux. The Gauss law of $\alpha$, where $a=\alpha\mathbf{1}_4$, yields $36\rho_{\sigma,v}+36\left<\mathrm{d}\alpha\right>/(2\pi)=0$, fixing the effective filling of each $\phi_{\sigma,v}$ at $\nu_{\sigma,v}=-1$.

In systems with threefold rotational symmetry, such as multilayer graphene, the nine valleys are naturally organized into three triplets related by the three-fold rotation. Within one triplet, the state formed by $\phi_{\sigma,v}$ can be described by a parton construction $\phi_{\sigma,v}=\xi_\sigma\eta_v$ for $v=1,2,3$. The fermionic partons $\xi_\sigma$ and $\eta_v$ carry the $U(4)$ gauge and the valley indices, respectively. As $\nu_{\sigma,v}=-1$, a simple choice is to place $\xi$ and $\eta$ in filled Chern bands with Chern numbers $C=1$ and $C=-1$, respectively. Upon integrating out the gapped partons, each triplet contributes a $U(4)_{1,-12}$ CS response to the gauge field $a$. Applying the same construction to all three triplets gives a total $U(4)_{3,-36}$ response from the dopants described by $\mathcal{L}[\phi,a]$ in \Eq{eq:Jain}. Adding this $U(4)_{3,-36}$ CS term to the $U(4)_{1,36}$ part describing the Jain-$4/9$ parent state, we recover $\mathcal{L}_\mathrm{Jain~4/9}^\mathrm{doped}[A]=\mathcal{L}_{2ne}^{\mathrm{CS}}[A]$ at $n=2$, establishing the $U(4)_{4,0}$ topological charge-$4e$ superconductor discussed above. The recent observation of a fractional quantum anomalous Hall insulator at $\nu=4/9$ in rhombohedral multilayer graphene~\cite{Lu2024FQAH} opens opportunities to explore this topological charge-$4e$ superconductor in experimentally accessible platforms. Superconductivity upon doping chiral spin liquids and Chern insulators has been established in microscopic numerical studies~\cite{Divic2025,Kuhlenkamp2025}, although the specific state $U(4)_{4,0}$ has not yet been examined.


{\it Higgsing the Read--Rezayi-type quantum Hall states.---}
A Moore--Read quantum Hall state can be viewed as the $(p+ip)$-wave pairing state of composite fermions coupled to a dynamical $U(1)$ gauge field~\cite{Moore1991}. By Higgsing this dynamical gauge field and identifying it with the background electromagnetic field, the composite fermion is converted into a physical electron, and the Moore--Read state to a $(p+ip)$-wave topological superconductor~\cite{Read2000}. At the wavefunction level, this Higgsing procedure is implemented by removing the Laughlin--Jastrow factor $\prod_{i<j}(z_i-z_j)^2$~\cite{Son2015}, leaving only the Pfaffian part $\mathrm{Pf}[(z_i-z_j)^{-1}]$ encoding the $(p+ip)$-wave pairing. The $2n$-cluster Read--Rezayi states are generalizations of the Moore--Read state, whose wavefunctions incorporate cluster structures formed by $2n$ particles and recover the Moore--Read state at $n=1$~\cite{Read1999}. 

Inspired by the $(p+ip)$-wave topological superconductor arising from Higgsing a Moore--Read state, a straightforward attempt to construct a topological charge-$2ne$ superconductor by Higgsing a $2n$-cluster Read--Rezayi state is to remove the Laughlin--Jastrow factor in its wavefunction $\Psi_{\mathrm{RR}\,2n}(\{z\})$~\cite{Read1999}:
$$\tilde{\Phi}(\{z\})=\Psi_{\mathrm{RR}\,2n}(\{z\})\prod_{k<l}(z_k-z_l)^{-1-\frac{1}{n}}\propto\left<\prod_i\chi_1(z_i)\right>,$$
where $\chi_1(z)$ denotes the elementary parafermion field in the edge $\mathbb{Z}_{2n}$ parafermion CFT. However, $\tilde{\Phi}(\{z\})$ is not a legal electron wavefunction, as it is not antisymmetric under electron exchanging. This can be seen from the conformal weight $h_\chi=1-1/(2n)$ of the parafermion field, which is not a half-odd integer unless $n=1$.

Several routes can compensate for the topological spin of $\chi_1(z)$ and construct a legal electron wavefunction. The most direct route is to retain a residual Laughlin--Jastrow factor; however, such a factor leads to a residual Hall conductivity, and hence the wavefunction describes a charge-gapped state instead of a superconducting state~\cite{supp}. Without introducing extra Laughlin--Jastrow factors, the two simplest routes are to replace $\chi_1(z)$ by the product of $n$ copies, $\prod_{q=1}^{n}\chi_1^{(q)}(z)$, or by the non-elementary parafermion field $\chi_n(z)$ in the $\mathbb Z_{2n^2}$ parafermion theory. Both choices have conformal weight $n-1/2$ and fusion order $2n$, satisfying the electron statistics and clustering requirements. 

Replacing $\chi_1(z)$ by the product of its $n$-copy leads to wavefunction $\Phi_{2ne}^\mathrm{PF}(\{z\})\propto\prod_{q=1}^n\big<\prod_i \chi_1^{(q)}(z_i)\big>$. The edge electron operator is identified as $\psi(z)=\prod_{q=1}^n\chi_1^{(q)}(z)$, the product of elementary parafermion fields in $n$ copies of the $\mathbb{Z}_{2n}$ parafermion theory. It is worth noting that this wavefunction has a clear relationship to the $2n$-cluster Read--Rezayi state,
\beq
\Phi_{2ne}^\mathrm{PF}(\{z\})=\big[\Psi_{\mathrm{RR}\,2n}(\{z\})\big]^n\prod_{k<l}(z_k-z_l)^{-n-1},\label{eq:2neRR}
\eeq
which remains a faithful wavefunction for particle number divisible by $2n$ up to non-holomorphic factors. In fact, it can be constructed by a parton decomposition of the electron, $\psi=\prod_{q=1}^n\xi_q\prod_{q'=1}^{n+1}\eta_{q'}$, where the fermionic partons $\xi_q$ and $\eta_{q'}$ form the $2n$-cluster Read--Rezayi state and the $C=-1$ Chern insulator, respectively. The TQFT describing this state is a CS theory with $[U(2)^{\otimes n}]/\mathbb{Z}_{2n}$ gauge group and chiral central charge $c_-=(2n^2+1)/(n+1)$, which exhibits non-Abelian topological order (see End Matter). For $n=2$, the topological charge-$4e$ superconductor hosts dynamical anyons with integer quantum dimensions, $h/(4e)$ vortices with quantum dimension $\sqrt{3}$ or $2\sqrt{3}$, and $h/(2e)$ vortices all built from attaching a dynamical anyon and/or an electron to a fixed Abelian vortex~\cite{supp}.

Replacing $\chi_1(z)$ by $\chi_n(z)$ in the $\mathbb{Z}_{2n^2}$ parafermion theory gives the wavefunction $\Phi_{2ne}^\mathrm{PF}(\{z\})\propto\big<\prod_i \chi_n(z_i)\big>$, where the edge electron operator is identified with $\chi_n(z)$. Such a wavefunction does not exhibit a simple relation to Read--Rezayi states; however, it has a clear TQFT description. A $\mathbb{Z}_{2n^2}$ parafermion theory can be described by the $U(2)_{2n^2,-4n^2}$ CS theory~\cite{Seiberg2016,Ma2020,Zou2020,Shi2025}, in which identifying the electron with $\chi_n(z)$ yields a $U(2)_{2n^2,0}$ CS theory with chiral central charge $c_-=3n^2/(n^2+1)$~\cite{supp},
\beq
\mathcal{L}_{2ne}^\mathrm{PF}[A]&=&\frac{2n^2}{4\pi}\trace{b\mathrm{d}b+\frac{2}{3}b^3}-\frac{n^2}{4\pi}(\mathrm{tr}\,b)\mathrm{d}(\mathrm{tr}\,b)\nn\\
&&-\frac{n}{2\pi}(\mathrm{tr}\,b)\mathrm{d}A,\label{eq:LPF}
\eeq
where $b$ is a $U(2)$ gauge field. This state also exhibits non-Abelian topological order, yet different from that described by wavefunction \Eq{eq:2neRR}. For $n=2$, the topological charge-$4e$ superconductor contains two inequivalent Fibonacci anyons with topological spin $3/5$~\footnote{For a fermionic theory the topological spin should be defined mod $1/2$. We nevertheless keep $3/5$ here to emphasize the identification of the bulk topological order with $(F_4)_1\otimes (F_4)_1$.} and quantum dimension $(1\!+\!\sqrt{5})/2$, suggesting an $(F_4)_1\otimes (F_4)_1$ bulk topological order. Its $h/(4e)$ vortices have quantum dimensions $\sqrt{10\!+\!2\sqrt{5}}/2$ and $\sqrt{5\!+\!2\sqrt{5}}$, and $h/(2e)$ vortices are again obtained by attaching a bulk anyon and/or an electron to a trivial Abelian vortex~\cite{supp}.

Substitutions of $\chi_1(z)$ beyond the above two minimal ones exist, leading to a family of topological charge-$2ne$ superconductors inspired by the $2n$-cluster Read--Rezayi state at a fixed $n$. This family can be generically constructed from a $\bigotimes_q\mathbb{Z}_{2nm_q}$ parafermion theory with $\sum_qm_q=n$~\cite{supp}. The edge electron operator is identified as $\psi(z)=\prod_{q}\chi_{m_q}(z)$ with conformal weight $n-1/2$, where $\chi_{m_q}$ is the $m_q$-th parafermion field in the $\mathbb{Z}_{2nm_q}$ parafermion theory. Taking $m_q=1$ for all $q$, or a single $m_1=n$ recovers the two minimal constructions above.

{\it Weak, intermediate, and strong clustering.---}
The two constructions differ in whether intrinsic non-Abelian structure persists in $h/(2e)$ vortices. In the $U(4)_{4,0}$ state, both $h/(4e)$ and $h/(2e)$ vortices are intrinsically non-Abelian, which cannot be attributed to attached dynamical anyons. By contrast, the charge-$4e$ states constructed from parafermion theories admit an Abelian $h/(2e)$ reference vortex, from which all $h/(2e)$ vortices are obtained by attaching dynamical anyons and/or electrons, although their elementary $h/(4e)$ vortices remain intrinsically non-Abelian.

\begin{figure}[htbp]
  \centering
  \includegraphics[width=0.8\linewidth]{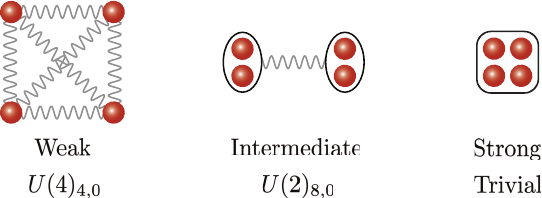}
  \caption{Weak, intermediate, and strong quartetting. Left: the weak-quartetting state described by the $U(4)_{4,0}$ CS theory, in which both $h/(4e)$ and $h/(2e)$ vortices are intrinsically non-Abelian. Middle: the intermediate-quartetting state described by the $U(2)_{8,0}$ CS theory, a topological pair superfluid of structureless Cooper pairs, in which $h/(4e)$ vortices remain intrinsically non-Abelian while $h/(2e)$ vortices are not. Right: the strong-quartetting state, a conventional BEC of structureless quartets without intrinsic topological order, in which all vortices are trivial.}
  \label{fig:wis}
\end{figure}

The latter states have a natural physical interpretation as topological pair superfluids of Cooper pairs. In this picture, tightly bound electron pairs act as structureless charge-$2e$ bosons, and pairs of these bosons condense while individual Cooper pairs remain gapped. This is already visible in the field theory \Eq{eq:LPF}: for $n=2$, its last term can be written as $-(\mathrm{tr}\,b)\,\mathrm{d}(2A)/(2\pi)$, so that each $2\pi$ flux of $\mathrm{tr}\,b$ carries charge $2e$ and $\mathrm{d}(\mathrm{tr}\,b)/(2\pi)$ is identified with the density of structureless charge-$2e$ Cooper pairs. The nontrivial topology resides in the collective paired state of these charge-$2e$ bosons, rather than in their internal electronic structure. An $h/(4e)$ vortex imposes a $\pi$ phase on a Cooper pair transported around it. In contrast, an $h/(2e)$ vortex produces a full $2\pi$ phase, equivalent to no phase rotation of these effective bosons. Its topological content is therefore the same as that of the vortex-free bulk, which amounts to an Abelian reference vortex bound to bulk anyons and/or electrons. Any non-Abelian degrees of freedom belong to these bound excitations, instead of the vortex itself. Such a topological pair superfluid of structureless Cooper pairs, represented by $U(2)_{8,0}$, realizes an \emph{intermediate-quartetting} topological charge-$4e$ superconductor. It lies between the \emph{weak-quartetting} $U(4)_{4,0}$ state, whose intrinsically non-Abelian $h/(2e)$ vortices obstruct a description in terms of structureless Cooper pairs, and the \emph{strong-quartetting} state, a conventional Bose--Einstein condensate (BEC) of tightly bound quartets without intrinsic topological order, whose $h/(4e)$ and $h/(2e)$ vortices are all Abelian. This extends the weak- and strong-pairing dichotomy of the $(p+ip)$-wave charge-$2e$ superconductor established by Read and Green~\cite{Read2000}, with the clustering level diagnosed by the vortex sectors carrying flux greater than $h/(4e)$. See Fig.~\ref{fig:wis} for illustrations.

The weak--intermediate--strong quartetting transitions can be described by introducing two complex adjoint Higgs fields $X,\,Y$ coupled to the $U(4)$ gauge field $a$ in \Eq{eq:LCS} with $n=2$, leading to the Higgs field theory,
\begin{equation}
    \mathcal{L}_\mathrm{H}[A]=\mathcal{L}_{4e}^\mathrm{CS}[A]+\mathrm{tr}\,|D_aX|^2+\mathrm{tr}\,|D_aY|^2+V(X,Y),\label{eq:Higgs}
\end{equation}
where $D_{a,\mu}X=\partial_\mu X-i[a_\mu,X]$, and similarly
for $Y$. The potential $V(X,Y)$ is chosen such that $\left<X\right>=v_X\tau^+\otimes\mathbf{1}_2$ whenever $X$ condenses and $\left<Y\right>=v_Y\mathbf{1}_2\otimes\tau^+$ whenever $Y$ condenses. Here $\tau^+=(\tau^x+i\tau^y)/2$ with Pauli matrices $\tau^{x,y}$. Tuning the potential yields phases with both Higgs fields gapped, only $X$ or $Y$ condensed, or both fields condensed (see End Matter). When both fields are gapped, \Eq{eq:Higgs} describes the weak-quartetting $U(4)_{4,0}$ state. Condensing only $X$ or $Y$ Higgses the gauge group to a diagonal $U(2)$, with the surviving gauge field embedded as $a=\mathbf{1}_2\otimes b$ or $a=b\otimes\mathbf{1}_2$, respectively. In either case, the CS terms reduce to the $U(2)_{8,0}$ theory in \Eq{eq:LPF}, identifying an intermediate-quartetting state. Condensing the remaining field further Higgses $U(2)$ to its $U(1)$ center, $b=\alpha\mathbf{1}_2$, which leaves a topological response $-4/(2\pi)\,\alpha\mathrm{d}A$ and suggests the topologically trivial strong-quartetting state.

For the charge-$4e$ superconductors, the two single-field condensation routes yield the same TQFT for intermediate quartetting. For general charge-$2ne$ superconductors with $n>2$, they lead to distinct intermediate-clustering states, which can arise from either $n$-clustering of Cooper pairs or pairing of $n$-clusters. Starting from the $U(2n)_{2n,0}$ weak-clustering state, condensing only $X$ yields the $U(n)_{4n,0}$ state, while condensing only $Y$ instead yields the Read--Rezayi-inspired $U(2)_{2n^2,0}$ state. The $U(n)_{4n,0}$ state is a topological $n$-cluster superfluid of tightly bound Cooper pairs, while individual Cooper pairs remain gapped. Its $h/(2e)$ vortices are formed by attaching dynamical anyons to a trivial Abelian vortex and are therefore not intrinsically non-Abelian. The $U(2)_{2n^2,0}$ state is a topological pair superfluid of tightly bound $n$-electron clusters, while these individual $n$-clusters remain gapped. Its $ph/(ne)$ vortices for all $p=1,2,\cdots,n\!-\!1$ are not intrinsically non-Abelian. Further condensing the remaining Higgs field in either intermediate-clustering state yields the topologically trivial strong-clustering state.

The weak--intermediate--strong clustering hierarchy exhibits an even--odd effect in $n$. For even $n$, embedding $U(2)$ into $U(n)$ as $n/2$ identical two-dimensional blocks gives a further Higgs route from $U(n)_{4n,0}$ to $U(2)_{2n^2,0}$. Physically, the tightly bound
$n$-electron clusters can themselves be assembled from Cooper
pairs. For odd $n$, however, this embedding is unavailable, and a direct continuous transition between the two intermediate-clustering states is unlikely. Moreover, the $n$-clusters composed of an odd number of electrons are fermionic in the $U(2)_{2n^2,0}$ state, suggesting the state is an intrinsically fermionic symmetry-enriched topological phase. For example, a charge-$6e$ state can organize as paired fermionic trions, described by the $U(2)_{18,0}$ fermionic CS theory, or as a three-particle condensate of Cooper pairs, described by the bosonic $U(3)_{12,0}$.

{\it Discussion.---}
We have developed a wavefunction and field-theory construction of topological charge-$2ne$ superconductors and characterized their edge CFTs, bulk anyon content, and superconducting vortices. Generalizing the Read--Green wavefunction and adapting Read--Rezayi-type parafermionic structures produce two distinct families of topological charge-$2ne$ superconductors, while doping Jain states supplies a more microscopic route to the former. Higgs theories connect these two families and organize them into a weak--intermediate--strong clustering hierarchy at fixed condensate charge.

The clustering hierarchy distinguishes how topology is organized in the charge-$2ne$ superconductors. For charge-$4e$, weak quartetting retains intrinsically non-Abelian $h/(2e)$ vortices and thereby cannot be described as a pair superfluid of structureless Cooper pairs. Intermediate quartetting admits such a description, while strong clustering removes intrinsic topological order. For higher charges, pairing $n$-electron clusters and clustering Cooper pairs yield inequivalent intermediate organizations, with constituent statistics distinguishing even and odd $n$.

The constructions identify candidate phases and their topological properties, while determining microscopic mechanisms favoring them remains an important future step~\cite{Wang20266e}. Doping the Jain fractional quantum anomalous Hall insulators provides a setting for investigating this energetic question in realistic microscopic models of experimental platforms, such as rhombohedral multilayer graphene and moir\'e systems. Another direction is to determine the critical behavior of the Higgs theories, including when the weak--intermediate--strong clustering transitions are genuinely continuous. The non-Abelian anyons and vortices also provide targets for future interferometric probes.

{\it Acknowledgment.---} 
We acknowledge Xiao-Gang Wen, Ashvin Vishwanath, Eslam Khalaf, Srinivas Raghu, Yi Zhang, Zhehao Dai, Zhaoyu Han, Taige Wang, Pavel Nosov, Clemens Kuhlenkamp, and especially Ya-Hui Zhang for helpful discussions. ZQG is supported by the Berkeley graduate program. YQW is supported by the JQI postdoctoral fellowship at the University of Maryland. CW is supported by the National Natural Science Foundation of China under Grant No. 12234016 and the New Cornerstone Science Foundation. We used ChatGPT 5.6 Pro and 6 Astra to assist with the derivations in the End Matter and Supplemental Material and to improve the clarity of parts of the text.

{\it Note added.---} Upon finishing this manuscript, we became aware of a recent study~\cite{shi2025nonabelian} on topological superconductors, including charge-$2ne$ ones, from a different point of view.

\bibliographystyle{apsrev4-2}

\bibliography{4e.bib}


\begin{center}
    {\bf End Matter}
\end{center}

{\it Field theory for the wavefunction \Eq{eq:2neRR}.}---
Wavefunction \Eq{eq:2neRR} can be constructed by a parton decomposition of the electron, $\psi=\prod_{q=1}^n\xi_q\prod_{q'=1}^{n+1}\eta_{q'}$, where fermionic parton $\xi_q$ and $\eta_{q'}$ forms the $2n$-cluster Read--Rezayi state and the $C=-1$ Chern insulator, respectively. The gauge constraints identify their currents $j^\mu_{\xi_1}=\cdots=j^\mu_{\xi_n}=j^\mu_{\eta_1}=\cdots=j^\mu_{\eta_{n+1}}$, which can be represented by a single dual gauge field $2\pi j^\mu=\epsilon^{\mu\nu\lambda}\partial_\nu b_\lambda$. Consequently, the TQFT describing the $2n$-cluster Read--Rezayi states formed by a single $\xi_q$ is,
\beq
\mathcal{L}_{\xi_q}&=&\frac{2n}{4\pi}\trace{a_q\mathrm{d}a_q+\frac{2}{3}a_q^3}-\frac{2n}{4\pi}(\mathrm{tr}\,a_q)\mathrm{d}(\mathrm{tr}\,a_q)\nn\\
&&+\frac{1}{2\pi}b\mathrm{d}(\mathrm{tr}\,a_q)+\frac{1}{4\pi}b\mathrm{d}b,
\eeq
where $a_q$ ($q=1,2,\cdots,n$) is a $U(2)$ gauge field. On the other hand, each $\eta_{q'}$ in a $C=-1$ Chern insulator contributes $-1/(4\pi)\,b\mathrm{d}b$, while the coupling to the electromagnetic field is $-1/(2\pi)\,b\mathrm{d}A$. Combining these contributions and integrating out $b$, we obtaine the resulting theory describing the topological charge-$2ne$ superconductor,
\beq
&&\mathcal{L}_{2ne}^\mathrm{PF}[A]=\nn\\
&&\frac{2n}{4\pi}\sum_{q}\trace{a_q\mathrm{d}a_q+\frac{2}{3}a_q^3}-\frac{1}{4\pi}\bigg(\sum_{q} \mathrm{tr}\,a_q\bigg)\mathrm{d}\bigg(\sum_{q} \mathrm{tr}\,a_q\bigg)\nn\\
&&-\frac{2}{4\pi}\sum_{q<q^\prime} \trace{a_q-a_{q^\prime}}\mathrm{d}\,\trace{a_q-a_{q^\prime}}-\frac{1}{2\pi}\sum_{q}(\mathrm{tr}\,a_q)\mathrm{d}A,\nn\\
\label{eq:LPF00}
\eeq
which exhibits a non-Abelian topological order for $n>1$ and chiral central charge $c_-=(2n^2+1)/(n+1)$. It reduces to the $(p+ip)$-wave topological superconductor for $n=1$ upon stacking with an invertible fermionic phase. The anyon content for the $n=2$ case are summarized in Table~\ref{tab:1}.

\begin{table}[!htbp]
\begin{tabular}{ccccccccc}
\hline\hline
Anyon type & $1$ & $\beta$ & $\epsilon$ & $\epsilon^\prime$ & $\sigma_+$ & $\sigma_-$ & $\lambda$ & $\lambda^\prime$  \\
\hline
Topo. spin & $0$ & $0$ & $1/3$ & $1/3$ & $1/6$ & $1/6$ & $1/8$ & $1/8$ \\
Quantum dim. & $1$ & $1$ & $2$ & $2$ & $2$ & $2$ & $3$ & $3$\\
\hline\hline
\end{tabular}
\caption{anyon content for topological charge-$4e$ superconductor described by wavefunction \Eq{eq:2neRR}.}\label{tab:1}
\end{table}

{\it Higgs potential in \Eq{eq:Higgs}.}---
We choose the Higgs potential $V(X,Y)$ in \Eq{eq:Higgs} to be,
\beq
V(X,Y)&=&r_X\,\trace{X^\dagger X}+r_Y\,\trace{Y^\dagger Y}\nn\\
&&+\,u_1\,\trace{\{X,X^\dagger\}^2+\{Y,Y^\dagger\}^2}\nn\\
&&+\,u_2\,\trace{(X^2)^\dagger X^2+(Y^2)^\dagger Y^2}\nn\\
&&+\,u_3\,\trace{[X,Y]^\dagger[X,Y]+[X,Y^\dagger]^\dagger[X,Y^\dagger]},\nn\\
\eeq
where $u_{1,2,3}>0$. Its global minima are attained at $\left<X\right>=0$ for $r_X>0$ and $\left<X\right>=v_X\tau^+\otimes\mathbf{1}_2$ for $r_X<0$, and similarly, $\left<Y\right>=0$ for $r_Y>0$ and $\left<Y\right>=v_Y\mathbf{1}_2\otimes\tau^+$ for $r_Y<0$. Here $v_{X,Y}=\sqrt{-r_{X,Y}/(4u_1)}$. The $u_2$-term and $u_3$-term penalize deviations from $X^2=Y^2=0$ and $[X,Y]=[X,Y^\dagger]=0$, respectively, which ensure the global minima above.

\clearpage


\onecolumngrid

\vspace{0.3cm}

\supplementarysection
 
\begin{center}
\Large{\bf Supplemental Material for ``Topological Charge-$2ne$ Superconductors"}
\end{center}

\section{Derivations of bulk TQFTs}

\subsection{Bulk TQFTs of weak clustering states}

Consider the $SU(2n)_{2n}$ CS theory with a charge $U(1)_0$ piece:
\beq
\mathcal{L}_0[A]=-\frac{1}{2\pi}b\mathrm{d}A+
\frac{2n}{4\pi}\trace{a\mathrm{d}a+\frac{2}{3}a^3},
\eeq
where $b$ is a $U(1)$ gauge field and $a$ is the $SU(2n)$ gauge field. Hence $\trace{a}=0$. Gauging the $\mathbb{Z}_{2n}$ center amounts to the field redefinition $a\mapsto a-b\mathbf{1}_{2n}/(2n)$, which results in
\beq
\mathcal{L}_{2ne}^\mathrm{CS}[A]=\frac{1}{4\pi}b\mathrm{d}b-\frac{1}{2\pi}b\mathrm{d}A+\frac{2n}{4\pi}\trace{a\mathrm{d}a+\frac{2}{3}a^3}-\frac{1}{2\pi }b\mathrm{d}(\mathrm{tr}\,a).
\eeq
Integrating out $b$ gives rise to
\beq
\mathcal{L}_{2ne}^\mathrm{CS}[A]=-\frac{1}{4\pi} (\mathrm{tr}\,a)\mathrm{d}(\mathrm{tr}\,a)-\frac{1}{4\pi}A\mathrm{d}A+\frac{2n}{4\pi}\trace{a\mathrm{d}a+\frac{2}{3}a^3}-\frac{1}{2\pi }(\mathrm{tr}\,a)\mathrm{d}A,
\eeq
which produces a $U(2n)_{2n,0}$ CS theory in the main text. The corresponding chiral central charge is $c_-=2n^2- 1/2$. By setting $a=\tilde{a}\mathbf{1}_{2n}$, the mutual CS term between $\mathrm{tr}\,a$ and $A$ shows that the Lagrangian describes a charge-$2ne$ SC. We also note that a $\bigotimes_q SU(2nm_q)_{k_q}$ WZW CFT with $\sum_qk_qm_q=2n$ can also realize a topological charge-$2ne$ SC by condensing the electron identified as $\psi(z)=\prod_q\Lambda_{m_q}^{(q)}(z)$. However, the bulk TQFT does not admit a simple Lagrangian formulation. Such a state differs from the $U(2n)_{2n,0}$ state in correlation functions with multiples of $2n$ electrons.

\subsection{Bulk TQFTs of intermediate clustering states}

Consider the $\bigotimes_q\mathbb{Z}_{2nm_q}$ parafermion theory with $\sum_{q=1}^pm_q=n$:
\beq
\mathcal{L}_0[A]=-\frac{1}{2\pi}b\mathrm{d}A+
\sum_{q=1}^p\,\frac{2nm_q}{4\pi}\trace{a_q\mathrm{d}a_q+\frac{2}{3}a_q^3}-\frac{2nm_q}{4\pi}(\mathrm{tr}\,a_q)\mathrm{d}(\mathrm{tr}\,a_q).
\eeq
Gauging the $\mathbb{Z}_{2n}$ symmetry amounts to field variable substitutions $a_q\mapsto a_q-b\mathbf{1}_{2}/(2n)$, which results in
\beq
\mathcal{L}_{2ne}^\mathrm{PF}[A]=-\frac{1}{4\pi}b\mathrm{d}b-\frac{1}{2\pi}b\mathrm{d}A+\sum_{q=1}^p\,\frac{2nm_q}{4\pi}\trace{a_q\mathrm{d}a_q+\frac{2}{3}a_q^3}-\frac{2nm_q}{4\pi}(\mathrm{tr}\,a_q)\mathrm{d}(\mathrm{tr}\,a_q)+\frac{m_q}{2\pi}b\mathrm{d}(\mathrm{tr}\,a_q).\label{eq:S5}
\eeq
Integrating out $b$ yields
\beq
\mathcal{L}_{2ne}^\mathrm{PF}[A]&=&\sum_{q=1}^p\,\frac{2nm_q}{4\pi}\trace{a_q\mathrm{d}a_q+\frac{2}{3}a_q^3}-\frac{2nm_q}{4\pi}(\mathrm{tr}\,a_q)\mathrm{d}(\mathrm{tr}\,a_q)-\frac{m_q}{2\pi}(\mathrm{tr}\,a_q)\mathrm{d}A\nn\\
&&+\frac{1}{4\pi}\Bigg(\sum_{q}m_q\mathrm{tr}\,a_q\Bigg)\mathrm{d}\Bigg(\sum_{q}m_q\mathrm{tr}\,a_q\Bigg)+\frac{1}{4\pi}A\mathrm{d}A.\label{eq:LPFK}
\eeq
Its charge part reads
\beq
\mathcal{L}_\mathrm{charge}^\mathrm{PF}[A]&=&\frac{1}{4\pi}\Bigg(\sum_{q}2m_q\tilde{a}_q\Bigg)\mathrm{d}\Bigg(\sum_{q}2m_q\tilde{a}_q\Bigg)-\sum_{q=1}^p\Bigg(\frac{4nm_q}{4\pi}\tilde{a}_q\mathrm{d}\tilde{a}_q+\frac{2m_q}{2\pi}\tilde{a}_q\mathrm{d}A\Bigg)+\frac{1}{4\pi}A\mathrm{d}A,\nn\\
&=&\Bigg(\frac{1}{4\pi}\sum_{q,q^\prime<p}\tilde{a}_qK_{qq^\prime}\mathrm{d}\tilde{a}_{q^\prime}-\frac{1}{2\pi}\sum_{q<p}2m_q\tilde{a}_q\mathrm{d}A\Bigg)-\frac{2n}{2\pi}\tilde{a}_p\mathrm{d}A+\frac{1}{4\pi}A\mathrm{d}A,
\eeq
where $\tilde{a}_q$ is defined as $a_q=\tilde{a}_q\mathbf{1}_2$, and the $K$-matrix is $K_{qq^\prime}=4m_qm_{q^\prime}-4nm_q\delta_{qq^\prime}$ with $|\mathrm{det}(K)|=4(4n)^{p-2}\prod_{q=1}^pm_q$ contributing a chiral central charge $c_-=1-p$. The total chiral central charge of this state is
\beq
c_-=1-p+\sum_{q=1}^p\frac{3nm_q}{nm_q+1}=2p+1-\sum_{q=1}^p\frac{3}{nm_q+1},
\eeq
which is consistent with that of the $\bigotimes_q\mathbb{Z}_{2nm_q}$ parafermion CFT. In particular when $m=n$ the theory \Eq{eq:LPFK} is the $U(2)_{2n^2,0}$ CS theory \Eq{eq:LPF} in the main text with chiral central charge $c_-=3n^2/(n^2+1)$ and no residual Abelian topological order. When $m=1$, it recovers the state described by wavefunction \Eq{eq:2neRR} with chiral central charge $c_-=(2n^2+1)/(n+1)$, up to invertible phases with integer central charge that does not affect anyon and vortex properties.

\section{Doping the $\nu=2n/(2n-1)^2$ Jain state}

A $\nu=2n/(2n-1)^2$ Jain state can be described by a $K$-matrix $(K_\mathrm{Jain})_{ij}=\delta_{ij}+2n-2$, which also admits a $U(2n)_{1,2n(2n-1)^2}$ CS theory description. Doping this state with the elementary anyon in the fundamental representation $(\mathbf{2n},1)$ of $U(2n)$ is described by,
\beq
\mathcal{L}_\mathrm{Jain}^\mathrm{doped}[A]&=&\frac{1}{4\pi}\trace{a\mathrm{d}a+\frac{2}{3}a^3}+\frac{2n-2}{4\pi}(\mathrm{tr}\,a)\mathrm{d}(\mathrm{tr}\,a)-\frac{1}{2\pi}(\mathrm{tr}\,a)\mathrm{d}A+\mathcal{L}[\phi,a],
\eeq
where $\phi$ sources the elementary anyon with electric charge $e/(2n-1)^2$. On a lattice, the projective translation symmetries thereby satisfy $T_xT_y=e^{2\pi i/(2n-1)^2}T_yT_x$, which generates $(2n-1)^2$ valleys in the low-energy modes of $\phi$ denoted as $\phi_{\sigma,v}$, where $\sigma\in\mathbb{Z}_{2n}$ and $v\in \mathbb{Z}_{(2n-1)^2}$ label the $U(2n)$ gauge indices and valleys, respectively. An ansatz invariant under $U(2n)$ and the magnetic translations enforces equal densities of $\rho_{\phi_{\sigma,v}}=\rho$ for all $\sigma,\,v$ and no average non-Abelian gauge flux. The Gauss law for $\alpha$, the $U(1)$ part of $a$, suggests,
\beq
\rho+\frac{\left<\mathrm{d}\alpha\right>}{2\pi}=0,
\eeq
where $a=\alpha\mathbf{1}_{2n}$. As in the $U(4)$ case in the main text, we divide the $(2n-1)^2$ valleys into $(2n-1)$ groups, where each group contains $(2n-1)$ valleys, and introduce a parton construction of $\phi_{\sigma,v}$ in each group as $\phi_{\sigma,v}=\xi_\sigma\eta_v$. As a result, $\eta_v$ is coupled to a dynamical $U(1)$ gauge field $b$, while $\xi_\sigma$ is coupled to $a-b\mathbf{1}_{2n}$. They have densities $\rho_{\xi_\sigma}=(2n-1)\rho$ and $\rho_{\eta_v}=2n\rho$ such that $\sum_\sigma\rho_{\xi_\sigma}=\sum_v\rho_{\eta_v}=2n(2n-1)\rho$, and see fluxes $(\left<\mathrm{d}\alpha\right>-\left<\mathrm{d}b\right>)/(2\pi)$ and $\left<\mathrm{d}b\right>/(2\pi)$, respectively. The simplest mean-field ansatz $\left<\mathrm{d}b\right>=2n\left<\mathrm{d}\alpha\right>$ yields effective filling $\nu_{\xi_\sigma}=-\nu_{\eta_v}=1$, which suggests placing $\xi_\sigma$ and $\eta_v$ in $C=1$ and $C=-1$ Chern insulators, respectively. Integrating out $\xi_\sigma$ and $\eta_v$ gives rise to the topological response,
\beq
\mathcal{L}_{\xi,\eta}&=&\frac{1}{4\pi}\trace{(a-b\mathbf{1}_{2n})\mathrm{d}(a-b\mathbf{1}_{2n})+\frac{2}{3}(a-b\mathbf{1}_{2n})^3}-\frac{2n-1}{4\pi}b\mathrm{d}b\nn\\
&=&\frac{1}{4\pi}\trace{a\mathrm{d}a+\frac{2}{3}a^3}-\frac{1}{2\pi}(\mathrm{tr}\,a)\mathrm{d}b+\frac{1}{4\pi}b\mathrm{d}b\nn\\
&=&\frac{1}{4\pi}\trace{a\mathrm{d}a+\frac{2}{3}a^3}-\frac{1}{4\pi}(\mathrm{tr}\,a)\mathrm{d}(\mathrm{tr}\,a),
\eeq
which is a $U(2n)_{1,-2n(2n-1)}$ CS theory of $a$. Combining the $(2n-1)$ valley groups gives a $U(2n)_{2n-1,-2n(2n-1)^2}$ CS description of $\mathcal L[\phi,a]$. Adding this response to the $U(2n)_{1,2n(2n-1)^2}$ CS theory of the Jain parent yields the $U(2n)_{2n,0}$ topological charge-$2ne$ superconductor.

\section{Failure of Higgsing a single Read--Rezayi state}

A $2n$-cluster Read--Rezayi state can be described by the $\mathbb{Z}_{2n}$ parafermion CFT, complemented by a $U(1)$ Laughlin--Jastrow factor, as
\beq
\Psi_{\mathrm{RR}\,2n}(\{z\})\propto\left<\prod_i\chi_1(z_i)\right>\prod_{k<l}(z_k-z_l)^{\frac{1}{n}+1},
\eeq
where $\chi_1(z)$ is the elementary $\mathbb{Z}_{2n}$ parafermion field. To define a trial superconducting wavefunction for electrons, one can only remove an even power of the Laughlin--Jastrow factor. The simplest example is to remove a power-$2$ factor, which transforms the Moore--Read state to the $(p+ip)$-wave topological superconductor,
\beq
\Phi_\mathrm{trial}(\{z\})\propto\left<\prod_i\chi_1(z_i)\right>\prod_{k<l}(z_k-z_l)^{\frac{1}{n}-1}.\label{eq:wrong}
\eeq
However, for $n\neq 1$, the presence of the residual Laughlin--Jastrow factor is detrimental to superconductivity. It suggests that the bulk have a nonvanishing Hall conductivity and consequently a charge gap. This can be seen from the wavefunction itself. Through some algebra, the wavefunction in \Eq{eq:wrong} can be rewritten as,
\beq
\Phi_\mathrm{trial}(\{z\})&=&\mathcal{S}\left[\prod_{p=1}^{2n}\prod_{i<j}(w^{(p)}_i-w^{(p)}_j)^2\right]\prod_{k<l}(z_k-z_l)^{-1}\nn\\
&=&\mathcal{S}\left\{\prod_{q=1}^{n}\mathcal{S}\left[\prod_{i<j}(u^{(q)}_i-u^{(q)}_j)^2(v^{(q)}_i-v^{(q)}_j)^2\right] \right\}\prod_{k<l}(z_k-z_l)^{-1}\nn\\
&=&\mathcal{S}\left\{\prod_{q=1}^{n}\left[\Psi_\mathrm{MR}(x^{(q)})\prod_{i<j}(x^{(q)}_i-x^{(q)}_j)^{-1}\right]\right\}\prod_{k<l}(z_k-z_l)^{-1}\nn\\
&=&\mathcal{S}\left\{\prod_{q=1}^{n}\left[\Psi_\mathrm{MR}(x^{(q)})\prod_{i<j}(x^{(q)}_i-x^{(q)}_j)^{-1}\right]\left[\prod_{r=1}^{n}\prod_{i<j}(x^{(r)}_i-x^{(r)}_j)^{-1}\right]\sigma(\{x\})\prod_{s<t}(x^{(s)}-x^{(t)})^{-1}\right\}\nn\\
&=&\mathcal{S}\left\{\sigma(\{x\})\prod_{q=1}^{n}\left[\Psi_\mathrm{MR}(x^{(q)})\prod_{i<j}(x^{(q)}_i-x^{(q)}_j)^{-2}\right]\prod_{s<t}(x^{(s)}-x^{(t)})^{-1}\right\}\nn\\
&=&\mathcal{A}\left[\prod_{q=1}^{n}\mathrm{Pf}\left(\frac{1}{x^{(q)}_i-x^{(q)}_j}\right)\prod_{s<t}\frac{1}{x^{(s)}-x^{(t)}}\right],\label{eq:RRtrial}
\eeq
where the coordinates $z_k$ are divided into $n$ groups labeled by $x^{(q)}_i$ with $q=1,2,\cdots,n$, $\sigma(\{x\})$ denotes the sign of a given division, and $\mathcal{S}(\mathcal{A})$ denotes (anti-)symmetrization over all divisions. For $n=2$, the physical meaning of this wavefunction is then rather clear: starting from a bilayer $(p+ip)$-wave topological superconductor described by the Pfaffian in each layer, with the interlayer pairing, one hybridizes the two layers as reflected by the antisymmetrization of two groups of coordinates $\{x^{(1,2)}\}$. However, it fails to describe a superconducting state, because the wavefunction \Eq{eq:RRtrial} introduces too many unwanted interlayer pairings and entanglements, leading to a gapped ``charge liquid'' state rather than a charge-condensed superconducting state.

This can also be seen from field theory. On an open manifold, the edge electron is written as $\psi(z)=\chi_1(z)V_{n-1}(z)$, where $V_{n-1}(z)$ is a vertex operator in the $U(1)_{-n(n-1)}$ CFT corresponding to the $U(1)_{-n(n-1)}$ CS theory in the bulk. The $\mathbb{Z}_{2n}$ parafermion is described by a $U(2)_{2n,-4n}$ CS theory. According to bulk-edge correspondence, the bulk TQFT for wavefunction \Eq{eq:wrong} is
\beq
\mathcal{L}_{\mathrm{trial}}[A]=\frac{2n}{4\pi}\trace{a\mathrm{d}a+\frac{2}{3}a^3}-\frac{2n}{4\pi}(\mathrm{tr}\,a)\mathrm{d}(\mathrm{tr}\,a)-\frac{n(n-1)}{4\pi}b\mathrm{d}b-\frac{n}{2\pi}b\mathrm{d}A.\label{eq:HiggsL}
\eeq
where $a$ and $b$ are the $U(2)$ and $U(1)$ gauge fields, respectively. 
Identifying the electron as the transparent fermion amounts to gauging the $\mathbb{Z}_{2n}$ one-form symmetry it generates, which results in identification of the gauge fields $b\mapsto b/n$ and $a\mapsto a-b\mathbf{1}_2/(2n)$ such that the $\mathbb{Z}_{2n}$ one-form symmetry acts trivially. Upon this identification, the effective theory for the $2n$-cluster Read--Rezayi state differs from \Eq{eq:HiggsL} by a level-2 CS term of $b$, which is consistent with the symmetry breaking picture. Integrating out the new variable $b$ yields
\beq
\mathcal{L}_{\mathrm{trial}}[A]&=&\frac{2n}{4\pi}\trace{a\mathrm{d}a+\frac{2}{3}a^3}-\frac{2n-1}{4\pi}(\mathrm{tr}\,a)\mathrm{d}(\mathrm{tr}\,a)-\frac{1}{2\pi}(\mathrm{tr}\,a)\mathrm{d}A+\frac{1}{4\pi}A\mathrm{d}A.
\eeq
Setting $a=\tilde{a}\mathbf{1}_2$, it is clear that except for $n=1$, this state remains insulating with a residual Hall conductance.

\section{Derivation of anyon condensation}

\subsection{General recipe}

In a unitary modular tensor category (UMTC), an Abelian object $J$ with $J^k=1$ generates a $\mathbb{Z}_k$ 1-form symmetry. For another object $a$, the monodromy between $a$ and $J$ is the charge of $a$ under this $\mathbb{Z}_k$ 1-form symmetry,
\begin{equation}
Q_J(a)=h_J+h_a-h_{J\times a}\pmod 1,
\end{equation}
where $h$ stands for the conformal weight of a primary field or the topological spin of an anyon when the UMTC describes a (1+1)D CFT or a (2+1)D TQFT, respectively. The monodromy phase is
\begin{equation}
M_{J,a}=e^{2\pi i Q_J(a)}.
\end{equation}
In particular, if $J$ and $a$ are anyons in a TQFT, the monodromy phase describes their mutual statistics. For a product CFT or TQFT, monodromies add factorwise:
\begin{equation}
Q_{(J_1,J_2)}(a_1,a_2)=Q_{J_1}(a_1)+Q_{J_2}(a_2)\pmod 1.
\end{equation}

The concept of monodromy is crucial to the anyon condensation construction of edge CFT and bulk TQFT in this paper. In a UMTC, anyons with bosonic or fermionic self-statistics can be condensed. Condensing a bosonic anyon $J$ results in another UMTC, where only the anyons having trivial monodromy $M=1$ with $J$ survive, and they are organized into orbits of the 1-form symmetry generated by $J$. Those with $M\neq 1$ are confined. Condensing a fermionic $J$ amounts to stacking a trivial fermionic topological order (say, an atomic insulator) with only the vacuum and the transparent fermion $f$, and then condensing the composite object $(J,f)$. The result of the fermionic anyon condensation is a super-modular tensor category (sMTC) which describes fermionic CFT or fermionic TQFT. $J$ is identified with the transparent fermion, which braids trivially with all anyons. Anyons having monodromies $M=1,i,-1,-i$ with $J$ are organized into orbits and become dynamical anyons, $h/(4e)$ vortices, $h/(2e)$ vortices, and $3h/(4e)$ vortices, respectively.

\subsection{Operator content of the $U(4)_{4,0}$ topological charge-$4e$ superconductor}

The neutral sector of $U(4)_{4,0}$ fermionic TQFT is equivalent to $SU(4)_4/\mathbb{Z}_4$. The edge theory of the $SU(4)_4$ CS theory is the $SU(4)_4$ WZW CFT, with primaries labeled by the highest weights of the affine Lie algebra $su(4)_4$, $(\lambda_1,\lambda_2,\lambda_3)$ with $\lambda_1+\lambda_2+\lambda_3\le 4$. The physical electron is identified as the primary $(4,0,0)$ with conformal weight $3/2$, which generates the $\mathbb{Z}_4$ center symmetry of $SU(4)$, $(\lambda_1,\lambda_2,\lambda_3)\mapsto (4-\lambda_1-\lambda_2-\lambda_3,\lambda_1,\lambda_2)$. It can be shown that, among $SU(4)_4$ primaries, only 10 primaries organized into 3 orbits have trivial monodromy with respect to $(4,0,0)$ (transparent fermion omitted):
\bea
& 1\sim (0,0,0)\sim (4,0,0)\sim (0,4,0)\sim (0,0,4),\\
& \epsilon\sim (0,1,2)\sim (1,0,1)\sim (2,1,0)\sim (1,2,1),\\
& \sigma_\pm \sim (0,2,0)\sim (2,0,2).
\eea
The last short orbit splits into two primaries $\sigma_\pm$, with quantum dimension halved, $d_\pm=1+\sqrt{2}$. The topological spins of these anyons are inherited from their parents in $SU(4)_4$ CS theory. The $\sigma_\pm$ anyon satisfies the fusion rule $\sigma_\pm\times\sigma_\pm=1+\sigma_\pm+\sigma_\pm^f$ in the $SO(3)_3$ fermionic CS theory with $\sigma_\pm^f=\sigma_\pm\times f$, while $\epsilon=\sigma_+\times \sigma_-$ is their composite. The bulk TQFT is then $SO(3)_3\otimes SO(3)_3$. Static $h/(4e)$ vortices having $M=i$ monodromy with the transparent fermion are organized into 2 orbits,
\bea
& (1,0,0)\sim (3,1,0)\sim (0,3,1)\sim (0,0,3),\\
& (1,2,0)\sim (1,1,2)\sim (0,1,1)\sim (2,0,1),
\eea
with quantum dimensions $\sqrt{4\!+\!2\sqrt{2}}$ and $(1\!+\!\sqrt{2})\sqrt{4\!+\!2\sqrt{2}}$, respectively. Static $h/(2e)$ vortices with $M=-1$ monodromy to the transparent fermion are organized into 3 orbits,
\bea
& (0,0,2)\sim (2,0,0)\sim (2,2,0)\sim (0,2,2),\\
& (0,1,0)\sim (3,0,1)\sim (0,3,0)\sim (1,0,3),\\
& (1,1,1),
\eea
where the first two have quantum dimension $2+\sqrt{2}$, while the last one is split into two (not four as the transparent fermion has order two) distinct types of vortices with quantum dimension $2(1+\sqrt{2})$, respectively. Thus, all $h/(4e)$ and $h/(2e)$ vortices are intrinsically non-Abelian, suggesting a weak-quartetting topological charge-$4e$ superconductor.

\subsection{Operator content of the $[U(2)\otimes U(2)]/\mathbb{Z}_4$ topological charge-$4e$ superconductor}

As seen from the wavefunction, the edge theory of the $[U(2)\otimes U(2)]/\mathbb{Z}_4$ topological charge-$4e$ superconductor is the $[(\mathbb{Z}_4~\mathrm{PF})\times (\mathbb{Z}_4~\mathrm{PF})]/\mathbb{Z}_4$ fermionic CFT,
where the physical electron is the composite of two elementary parafermions $[(0,2),(0,2)]$. Gauging the $\mathbb{Z}_4$ symmetry converts the bosonic CFT into a fermionic one. The trivial monodromy condition requires dynamical anyons labeled by orbits of $[(l_1,m_1),(l_2,m_2)]$ to have $m_1+m_2=0$ (mod 4), leading to the orbits (transparent fermion omitted),
\bea
& 1\sim [(0,0),(0,0)]\sim [(0,2),(0,2)]\sim [(0,4),(0,4)]\sim [(0,6),(0,6)],\\
& \beta\sim [(0,0),(0,4)]\sim [(0,2),(0,6)]\sim [(0,4),(0,0)]\sim [(0,6),(0,2)],\\
&\epsilon\sim [(0,0),(2,0)]\sim [(0,2),(2,2)]\sim [(0,4),(2,0)]\sim [(0,6),(2,2)],\\
&\epsilon^\prime\sim [(2,0),(0,0)]\sim [(2,2),(0,2)]\sim [(2,0),(0,4)]\sim [(2,2),(0,6)],\\
&\lambda\sim [(1,1),(1,3)]\sim [(1,3),(1,5)]\sim [(1,5),(1,7)]\sim [(1,7),(1,1)],\\
&\lambda^\prime\sim [(1,3),(1,1)]\sim [(1,5),(1,3)]\sim [(1,7),(1,5)]\sim [(1,1),(1,7)],\\
&\sigma_\pm\sim [(2,0),(2,0)]\sim [(2,2),(2,2)],
\eea
where the last orbit with length 2 splits into two anyons $\sigma_\pm$. The topological spins, quantum dimensions, and fusion rules are inherited from the parent $(\mathbb{Z}_4~\mathrm{PF})\times (\mathbb{Z}_4~\mathrm{PF})$ theory. The $h/(4e)$ vortices having monodromy $M=i$ with $[(0,2),(0,2)]$ are,
\bea
&[(1,1),(0,0)]\sim [(1,3),(0,2)]\sim [(1,5),(0,4)]\sim [(1,7),(0,6)],\\
&[(0,0),(1,1)]\sim [(0,2),(1,3)]\sim [(0,4),(1,5)]\sim [(0,6),(1,7)],\\
&[(1,5),(0,0)]\sim [(1,7),(0,2)]\sim [(1,1),(0,4)]\sim [(1,3),(0,6)],\\
&[(0,0),(1,5)]\sim [(0,2),(1,7)]\sim [(0,4),(1,1)]\sim [(0,6),(1,3)],\\
&[(1,1),(2,0)]\sim [(1,3),(2,2)]\sim [(1,5),(2,0)]\sim [(1,7),(2,2)],\\
&[(2,0),(1,1)]\sim [(2,2),(1,3)]\sim [(2,0),(1,5)]\sim [(2,2),(1,7)],
\eea
where the first four and the last two have quantum dimension $\sqrt{3}$ and $2\sqrt{3}$, respectively. The $h/(2e)$ vortices having monodromy $M=-1$ with $[(0,2),(0,2)]$, however, are obtained by fusing the Abelian object $[(0,2),(0,0)]$ (or equivalently $[(0,0),(0,2)]$) to the dynamical anyons. Therefore, the $h/(2e)$ vortices are not intrinsically non-Abelian, where $[(0,2),(0,0)]$ plays the role of the reference Abelian vortex and fusing two identical reference vortices yields a dynamical anyon $\beta$. This indeed suggests an intermediate-quartetting topological charge-$4e$ superconductor.

\subsection{Operator content of the $U(2)_{8,0}$ topological charge-$4e$ superconductor}

The edge theory of the $U(2)_{8,0}$ topological charge-$4e$ superconductor is the $(\mathbb{Z}_8~\mathrm{PF})/\mathbb{Z}_4$ fermionic CFT, where the physical electron is identified with the order-$4$ parafermion field $(0,4)$. Gauging the $\mathbb{Z}_4$ symmetry converts the bosonic CFT to be a spin one. The trivial monodromy condition requires dynamical anyons labeled by orbits of $(l,m)$ to have $m=0$ (mod 4), leading to the orbits (transparent fermion omitted),
\bea
& 1\sim (0,0)\sim (0,4)\sim (0,8)\sim (0,12),\\
& \epsilon\sim (2,0)\sim (2,4)\sim (2,8)\sim (2,12),\\
& \varphi_\pm\sim (4,0)\sim (4,4),
\eea
where the last orbit with length 2 splits into two Fibonacci anyons $\varphi_\pm$ with topological spin $3/5$ and quantum dimension $(1+\sqrt{5})/2$, and $\epsilon=\varphi_+\times \varphi_-$. The bulk TQFT is then $(F_4)_1\otimes(F_4)_1$. The $h/(4e)$ vortices having monodromy $M=i$ with $(0,4)$ are organized into 2 orbits,
\bea
(1,1)\sim (1,5)\sim (1,9)\sim (1,13),\\
(3,1)\sim (3,5)\sim (3,9)\sim (3,13),
\eea
with quantum dimensions $\sqrt{10\!+\!2\sqrt{5}}/2$ and $\sqrt{5\!+\!2\sqrt{5}}$, respectively. The $h/(2e)$ vortices are obtained from fusing $(0,2)$ to the dynamical anyons, where $(0,2)$ is the Abelian reference vortex and fusing two identical $(0,2)$ yields an electron in the vacuum sector. These topological data are consistent with $U(2)_{8,0}$ CS theory upon stacking an invertible fermionic topological phase, and suggest an intermediate-quartetting topological charge-$4e$ superconductor described by $U(2)_{8,0}$.

\vfill 

\end{document}